\documentclass[10pt]{article}
\usepackage{amsmath}
\usepackage{graphicx}
\usepackage{color}
\usepackage{orcidlink}
\usepackage{caption}
\usepackage{subcaption}
\usepackage{hyperref}
\hypersetup{colorlinks=true, linkcolor=blue, citecolor=blue, urlcolor=blue}
\RequirePackage[numbers,sort&compress]{natbib}
\begin{document}
\baselineskip=18pt

\begin{center}
{\large {\bf Gravitational lensing in holonomy corrected spherically symmetric black holes with phantom global monopoles }}
\end{center}

\vspace{0.3cm}

\begin{center}
    {\bf Faizuddin Ahmed\orcidlink{0000-0003-2196-9622}}\footnote{\bf faizuddinahmed15@gmail.com}\\
    \vspace{0.2cm}
    {\it Department of Physics, University of Science \& Technology Meghalaya, Ri-Bhoi, Meghalaya, 793101, India}
\end{center}

\vspace{0.3cm}

\begin{abstract}
In this paper, we address a theoretical investigation of the gravitational lensing phenomenon within the space-time framework of a holonomy-corrected spherically symmetric black hole (BH), incorporating both ordinary and phantom global monopoles. Our focus lies on the analysis of null geodesics within this black hole background, examining the influence of ordinary and phantom global monopoles on the effective potential of null geodesics of the system. Afterwards, we derive analytical expressions for the deflection angle of photon light, considering weak field limit. The obtain expressions are presented up to the second order of the Loop Quantum Gravity parameter, enabling a thorough examination of the impact of ordinary and phantom global monopoles on the deflection angle.
\end{abstract}

\vspace{0.1cm}
{\bf keywords:} Modified theories of gravity; Gravitational lenses and luminous arcs; magnetic monopoles

\vspace{0.1cm}

\section{Introduction}

Physicists are actively engaged in tackling the singularity problem through the pursuit of a quantum theory of gravity, even though a comprehensive theory and experimental evidence on the quantum aspects of gravity are still lacking \cite{ss1}. An alternative approach focuses on describing certain phenomena at energy scales much lower than the Planck scale \cite{ss2,ss3,ss4}. These models not only offer insights into quantum gravity but also provide corrections from gravity to quantum electrodynamics \cite{ss5, ss6, ss7} and quantum chromodynamics (QCD) \cite{ss8}. Additionally, researchers explore the impact of quantum gravitation on dark matter \cite{ss9, ss10}. 

Among the various effective theories, Loop Quantum Gravity (LQG) has gained prominence. In cosmology, LQG has been utilized to construct cosmological models devoid of singularities \cite{ss11, ss12, ss13} and to provide explanations for phenomena such as the Big Bang \cite{ss14, ss15}. In the realm of black hole physics, the LQG formalism has been applied to derive a new expression for the Bekenstein-Hawking formula \cite{ss16}. In a different approach, researchers have employed the Schwarzschild metric as a background geometry within the classical polymerization framework, revealing that its energy-momentum tensor exhibits characteristics resembling dark energy \cite{ss17}. Numerous studies have explored LQG corrections to the Schwarzschild solution \cite{ss18, ss19}. Recently, a black hole model has been proposed by employing a canonical transformation and a linear combination of general relativity constraints, introducing an LQG correction to the Schwarzschild solution \cite{AAB1, AAB2}. This innovative approach showcases ongoing efforts to refine our understanding of gravity at both quantum and classical levels, with potential implications for resolving long-standing cosmological and astrophysical puzzles.

The line-element describing a holonomy corrected Schwarzschild black hole in the spherical coordinates $(t, r, \theta, \phi)$ is given by \cite{AAB1, AAB2, PRD1, PRD2}
\begin{equation}
ds^2=-\mathcal{A}(r)\,dt^2+\Big[\Big(1-\ell/r\Big)\,\mathcal{A}(r)\Big]^{-1}\,dr^2+r^2\,(d\theta^2+\sin^2 \theta\,d\phi^2), \label{a1}
\end{equation}
where 
\begin{equation}
    \mathcal{A}(r)=\Big(1-\frac{2\,M}{r}\Big).\label{a2}
\end{equation}
where $M$ is the mass of BH, $\ell$ is a new scale length defined by $\ell=\Big(\frac{2\,m\,\lambda^2}{1+\lambda^2}\Big)$ also called LQG parameter, where $\lambda$ is called the polymerization constant and provides the holonomy correction information. For $m > 0$, this solution is asymptotically flat and contains a globally hyperbolic black hole or white hole region with a minimal space-like hypersurface replacing the original singularity.

The line-element describing a spherically symmetric black hole with global monopoles (ordinary and phantom) in the spherical coordinates $(t, r, \theta, \phi)$ is given by \cite{SCJJ, ahep}
\begin{equation}
ds^2=-\mathcal{F}(r)\,dt^2+\frac{dr^2}{\mathcal{F}(r)}+r^2\,(d\theta^2+\sin^2 \theta\,d\phi^2), \label{a3}
\end{equation}
with the function
\begin{equation}
    \mathcal{F}(r)=\Big(1-8\,\pi\,\eta^2\,\xi-\frac{2\,M}{r}\Big),\label{a4}
\end{equation}
where $M$ is the mass, $\eta$ is the energy scale of symmetry breaking, and $\xi$ is the term describing kinetic energy of the BH. If $\xi=1$, it represents an ordinary global monopole originating from positive kinetic energy of scalar field \cite{AVV}. The properties of the global monopole in curved space-time, particularly its gravitational effects, were first studied by Barriola {\it et al.} \cite{AVV}. In their investigation, they demonstrated that the monopole exerts no gravitational force. The space surrounding and outside the monopole exhibits a solid deficit angle, which deflects all light passing near it. Harari {\it et al.} \cite{DH}, and later Shi {\it et al.} \cite{XS}, further examined the global monopole and have shown that the gravitational potential is actually repulsive. Additionally, Li {\it et al.} \cite{Li} studied the global monopole in asymptotically AdS/dS space-time and showed that the mass of the monopole might be positive in asymptotically dS space-time if the cosmological constant exceeds a critical value. They also demonstrated that the gravitational field of the global monopole could be attractive or repulsive depending on the value of the cosmological constant. The physics associated with the global monopole has been comprehensively studied in detail in \cite{ERBM}, and subsequently explored by numerous authors in the context of general relativity (see, for examples, \cite{HT,KAB,SLL}). If $\xi=-1$, the phantom global monopole is formed originating from negative kinetic energy of scalar field. Performing the following transformations 
\begin{equation}
    t \to (1-8\,\pi\,\eta^2\,\xi)^{1/2}\,t,\quad r \to (1-8\,\pi\,\eta^2\,\xi)^{1/2}\,r,\quad M \to (1-8\,\pi\,\eta^2\,\xi)^{3/2}\,M, \label{a5}
\end{equation}
into the metric (\ref{a3}), one can obtain the following line-element \cite{SCJJ}
\begin{equation}
ds^2=-\mathcal{A}(r)\,dt^2+\frac{dr^2}{\mathcal{A}(r)}+r^2\,(1-8\,\pi\,\eta^2\,\xi)\,(d\theta^2+\sin^2 \theta\,d\phi^2).\label{a6}
\end{equation}

Examining the above line-element, it is evident that for $\xi=1$, corresponding to an ordinary global monopole, there exists a deficit solid angle $(1-8\,\pi\,\eta^2)$ within the system $S_{+}$. Conversely, for $\xi=-1$, representing the phantom global monopole, the solid angle for the system $S_{-}$ becomes $(1+8\,\pi\,\eta^2)$, reflecting a surplus rather than a deficit. This observation implies that the topological properties of a space-time hosting a phantom global monopole differ from those associated with an ordinary global monopole. Global monopoles manifest as topological defects within the vacuum manifold, originating from phase transitions in the early universe \cite{AV}. Their formation is contingent upon gauge symmetry breaking with a suitable choice of a scalar field. It is noteworthy that the energy of global monopoles is concentrated near the monopole core within a confined region.

Gravitational lensing occurs when a massive object or concentration of matter, such as a galaxy cluster, induces a gravitational field that causes substantial curvature in space-time. As a result, the trajectory of light originating from a distant source deviates from a straight line, becoming curved on its journey towards the observer. This bending of light is a consequence of the gravitational field's impact on the path of photons, leading to changes in the apparent position and characteristics of the source as observed from Earth. Gravitational lensing is a pivotal and profound area of exploration in cosmology and gravitation, manifesting in both weak field scenarios, where light rays pass far from the source, and strong field situations, where light rays pass very close to massive objects. The study of gravitational lensing spans various space-time backgrounds, encompassing scenarios such as charged black holes within string theory \cite{aa1}, Schwrarzschild BH with global monopole charge in weak field limit \cite{ND} and strong field limit \cite{HC}, brane-world black holes \cite{aa2}, Reissner-Nordstrom black holes \cite{aa4}, space-times with naked singularities \cite{aa6, aa7, aa8, aa9}, Schwarzschild black holes \cite{aa10}, Kerr-Randers optical geometry \cite{aa13}, space-time within the framework of Eddington-inspired Born-Infeld theory with global monopoles \cite{aa32}, Kerr-MOG black holes \cite{aa34}, Simpson-Visser black-bounce space-time \cite{aa35}, rotating regular black holes \cite{aa37}, black-bounce space-time \cite{JRN}, black holes \cite{aa39}, Rindler modified Schwarzschild black holes \cite{bb1}, phantom black holes \cite{bb2}, wormholes \cite{RS,MA}, topologically charged Ellis-Bronnikov wormhole \cite{MA2}, topologically charged Eddington-inspired Born-Infeld space-time \cite{aa33}, Morris-Thorne wormholes with cosmic strings \cite{aa31}, and Eddington-inspired Born-Infeld gravity space-time with cosmic string \cite{kk}.

In this paper, we aim to investigate null geodesics of photon rays and gravitational lensing of spherically symmetric BHs (with ordinary and phantom global monopoles) under holonomy corrections. In fact, we show that the effective potential of the system gets modifications by phantom global monopole compared to that of with a ordinary global monopole. Moreover, we see that the deflection angle of photon rays in the weak field limit is influenced by the holonomy corrections in addition to the ordinary and phantom global monopoles. The format of this paper is as follows. In the next section, we introduce the black hole metric having both ordinary and global monopoles under holonomy corrections and study the behavior of null geodesics. Then, we derive the deflection angle of photon rays in the weak field approximation. In the last section, we summarize the results.

\section{Null geodesics of holonomy corrected black holes with phantom global monopoles and Gravitational lensing}

In this section, we study null geodesics within the context of a holonomy-corrected spherically symmetric black hole featuring global monopole, both of the ordinary and phantom varieties. Therefore, we begin this section by introducing the line-element of a holonomy-corrected spherically symmetric black hole in the spherical coordinates $(t, r, \theta, \phi)$ with ordinary and phantom global monopoles is given by
\begin{eqnarray}
   ds^2=-\mathcal{A}(r)\,dt^2+\Big[\Big(1-\ell/r\Big)\,\mathcal{A}(r)\Big]^{-1}\,dr^2+r^2\,(1-8\,\pi\,\eta^2\,\xi)\,(d\theta^2+\sin^2 \theta\,d\phi^2).\label{b1}
\end{eqnarray}
where the function $\mathcal{A}$ is given in Eq. (\ref{a2}). 

We see that in the limit $\xi \to 0$, black hole metric (\ref{b1}) reduces to a holonomy corrected Schwarzschild black hole solution (\ref{a1}). Furthermore, in the limit $\ell \to 0$, one will find black hole space-time with ordinary and phantom global monopoles (\ref{a6}). The Lagrangian of a system is defined by \cite{aa6, aa7, aa8, aa9, aa31}
\begin{equation}
\mathcal{L}=\frac{1}{2}\,g_{\mu\nu}\,\left(\frac{dx^{\mu}}{d\tau}\right)\,\left(\frac{dx^{\nu}}{d\tau}\right), \label{b2}
\end{equation}
where $\tau$ is the affine parameter of the curve, and $g_{\mu\nu}$ is the metric tensor. 

Using the line-element (\ref{b1}) and in the equatorial plane defined by $\theta=\frac{\pi}{2}$, we obtain
\begin{equation}
\mathcal{L}=\frac{1}{2}\,\Bigg[-\mathcal{A}(r)\,\dot{t}^2+\Big[\Big(1-\ell/r\Big)\,\mathcal{A}(r)\Big]^{-1}\,\dot{r}^2+r^2\,(1-8\,\pi\,\eta^2\,\xi)\,\dot{\phi}^2\Bigg].\label{b3}
\end{equation}
There are two constant of motions given by
\begin{eqnarray}
    &&-E=-\mathcal{A}(r)\,\dot{t} \Rightarrow \dot{t}=\frac{E}{\mathcal{A}(r)},\label{b4}\\
    &&L=(1-8\,\pi\,\eta^2\,\xi)\,r^2\,\dot{\phi}\Rightarrow \dot{\phi}=\frac{L}{(1-8\,\pi\,\eta^2\,\xi)\,r^2},\label{b5}
\end{eqnarray}
where $E$ is the conserved energy parameter, and $L$ is the conserved angular momentum. 

With these, the Lagrangian (\ref{b3}) for light-like or time-like geodesics becomes
\begin{equation}
    \Big(1-\ell/r\Big)^{-1}\,\Big(\frac{dr}{d\tau}\Big)^2+\Big(1-\frac{2\,M}{r}\Big)\,\Bigg[-\epsilon+\frac{L^2}{(1-8\,\pi\,\eta^2\,\xi)\,r^2}\Bigg]=E^2,\label{b6}
\end{equation}
where $\epsilon=0$ for null geodesics and $-1$ for time-like geodesics.

Eq. (\ref{b6}) can be seen as describing the dynamics of a classical particle of energy E subject to an effective potential given by
\begin{equation}
    V_{eff} (r)=\Big(1-\frac{2\,M}{r}\Big)\,\Bigg[-\epsilon+\frac{L^2}{(1-8\,\pi\,\eta^2\,\xi)\,r^2}\Bigg].\label{b7}
\end{equation}

\begin{center}
\begin{figure}[ht!]
\begin{centering}
\subfloat[$L=1$]{\centering{}\includegraphics[scale=0.72]{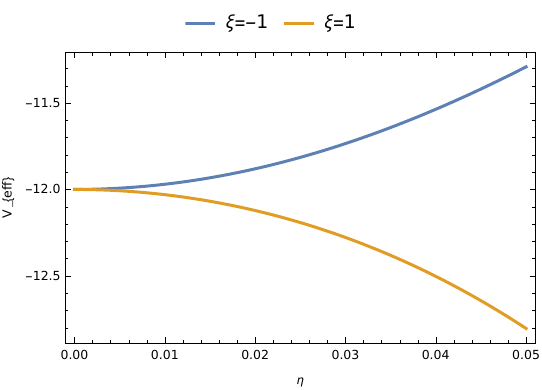}}\quad\quad\quad
\subfloat[$L=2$]{\centering{}\includegraphics[scale=0.72]{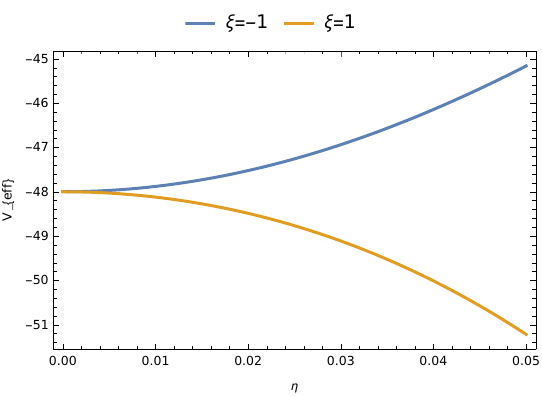}}\\
\subfloat[$L=3$]{\centering{}\includegraphics[scale=0.72]{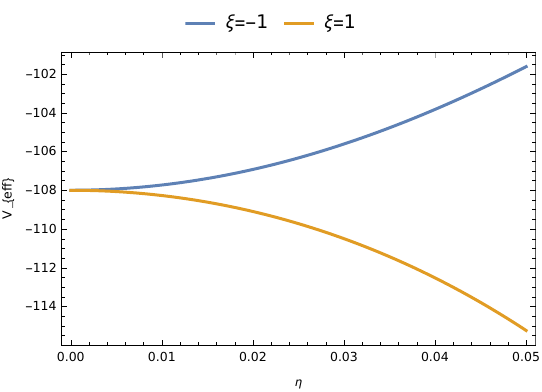}}\quad\quad\quad
\subfloat[$L=4$]{\centering{}\includegraphics[scale=0.72]{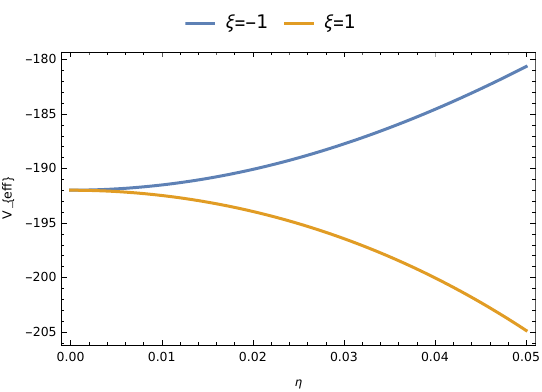}}
\caption{The effective potential for null geodesics. Here $M=1$ and $r=0.5$.}
\label{fig: 1}

\subfloat[$L=1$]{\centering{}\includegraphics[scale=0.72]{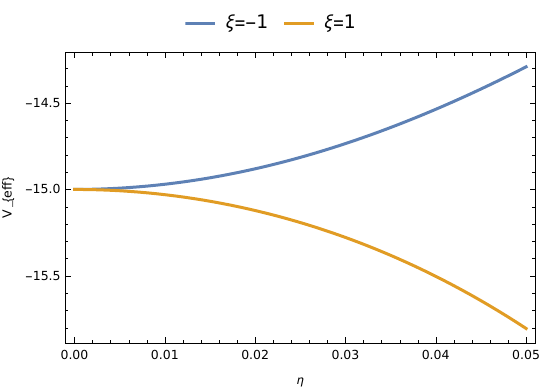}}\quad\quad\quad
\subfloat[$L=2$]{\centering{}\includegraphics[scale=0.72]{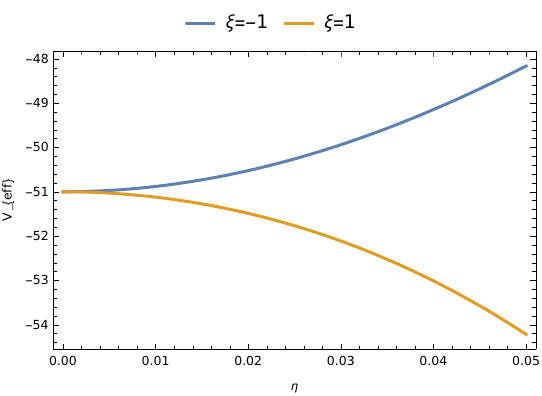}}\\
\subfloat[$L=3$]{\centering{}\includegraphics[scale=0.72]{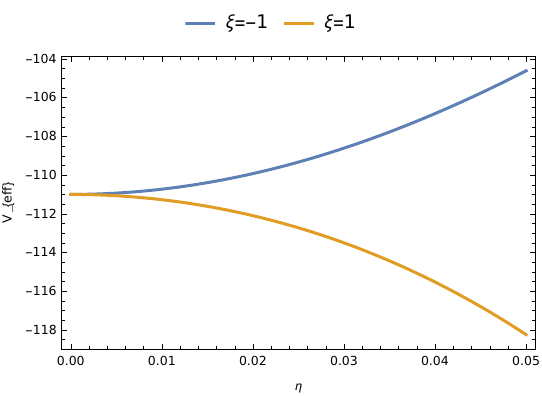}}\quad\quad\quad
\subfloat[$L=4$]{\centering{}\includegraphics[scale=0.72]{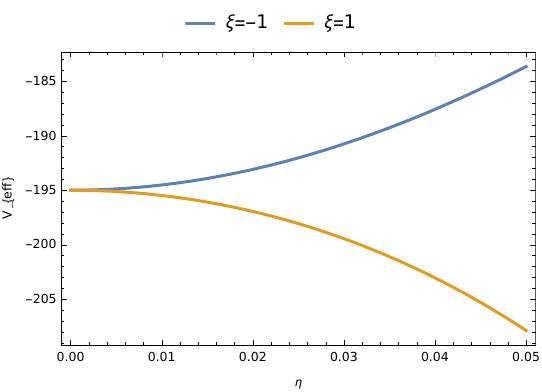}}
\caption{The effective potential for time-like geodesics. Here $M=1$ and $r=0.5$.}
\label{fig: 2}
\end{centering}
\end{figure}
\par\end{center}

In Figure 1, we have depicted the effective potential for null geodesics varying the energy scale $\eta$ for ordinary $\xi=1$ and phantom $\xi=-1$ global monopoles. In Figure 2, the effective potential for time-like geodesics is showcased once more for the same values of the conserved angular momentum. Throughout these figures, we have fixed $M=1$ unit and $r=1/2$ unit. We choose the conserved angular momentum $L=1, 2, 3, 4$, respectively in Figures \ref{fig: 1}--\ref{fig: 2}.

Now, we explore gravitational lensing in the weak field limit and examine the influences of ordinary and phantom global monopoles, incorporating holonomy corrections. Considering null geodesics where $\epsilon=0$, we deduce from Eq. (\ref{b6}) that:
\begin{equation}
    \dot{r}^2=E^2\,\Big(1-\ell/r\Big)\,\Bigg[1-\Big(1-\frac{2\,M}{r}\Big)\,\frac{\beta^2}{(1-8\,\pi\,\eta^2\,\xi)\,r^2}\Bigg],\label{c1}
\end{equation}
where we defined the impact parameter $\beta=L/E$ with the conserved energy parameter $E$ and the angular momentum $L$.

To analyze the spacetime lensing of a holonomy-corrected black hole with ordinary and phantom global monopoles, let's consider a photon originating from the asymptotically flat region and approaching the black hole at a radial distance $r_0$, referred to as the turning point. It is assumed that $r_0 > r_{ph}$, where $r_{ph}=3\,M$ represents the radius of the photon sphere in our context. After being influenced by the gravitational field of the black hole, the photon proceeds toward another asymptotically flat region. At the turning point, the energy $E$ is equal to the effective potential $V_{eff}(r_0)$, that is,
\begin{equation}
    \frac{\alpha^2}{\beta^2}=\frac{\Big(1-\frac{2\,M}{r_0}\Big)}{r^2_{0}},\quad \alpha^2=(1-8\,\pi\,\eta^2\,\xi),\label{c2}
\end{equation}
where $(\xi=\pm\,1)$.

Therefore, using Eqs. (\ref{b5}) and (\ref{c1}), one can define the following quantity
\begin{equation}
    \frac{\dot{r}}{\dot{\phi}}=\frac{dr}{d\phi}=r^2\,\sqrt{\alpha^2\,\Big(1-\ell/r\Big)}\,\sqrt{\frac{\alpha^2}{\beta^2}-\Big(1-\frac{2\,M}{r}\Big)\,\frac{1}{r^2}}.\label{c3}
\end{equation} 
By symmetry, the contributions to $\Delta\phi$ before and after the turning point are equal Hence, we can obtain the deflection angle
\begin{eqnarray}
\delta\phi=\Delta\phi-\pi, \label{c4}
\end{eqnarray}
where we have defined
\begin{eqnarray}
\Delta\phi=2\,\int^{\infty}_{r_0}\,\frac{dr}{r^2\,\sqrt{\alpha^2\,\Big(1-\ell/r\Big)\,\Bigg[\frac{\alpha^2}{\beta^2}-\Big(1-\frac{2\,M}{r}\Big)\,\frac{1}{r^2}\Bigg]}}.\label{c5}
\end{eqnarray}

Let’s introduce a new variable $u=\frac{1}{r}$, where $du=-\frac{dr}{r^2}$ in the above equation. One can see that at $r \to r_0$, we choose $u \to u_0$ and at $r \to \infty$, we have $u \to 0$. Then from Eq. (\ref{c2}), we can write $\frac{\alpha^2}{\beta^2}=u^2_{0}\,(1-2\,M\,u_0)$. Therefore, in terms of $u$, from Eq. (\ref{c5}), we obtain
\begin{eqnarray}
    \Delta\phi=2\,\int^{u_0}_{0}\,\frac{du}{\sqrt{\alpha^2\,(1-\ell\,u)\,\Big[u^2_{0}\,(1-2\,M\,u_0)-u^2+2\,M\,u^3\Big]}}. \label{c6}
\end{eqnarray}
In the weak field approximation, that is, assuming that the photon passes very far from the BH, we can take the approximation $M <<1$ and $\ell <<1$. Therefore, up to the
second order in $\ell$, from Eq. (\ref{c6}), we obtain
\begin{equation}
    \Delta\phi \simeq \frac{\pi}{\alpha}+\frac{4\,M}{\beta}+\frac{\ell}{\beta}+\frac{3\,\pi\,\ell^2\,\alpha}{16\,\beta^2}+\frac{M\,\ell\,(3\,\pi-4)\,\alpha}{4\,\beta^2}.\label{c7}
\end{equation}

Thus, the deflection angle from Eq. (\ref{c4}) and using (\ref{c7}) is given by
\begin{equation}
    \delta\phi \simeq \Big(\frac{1}{\sqrt{1-8\,\pi\,\eta^2\,\xi}}-1\Big)\,\pi+\frac{4\,M}{\beta}+\frac{\ell}{\beta}+\frac{3\,\pi\,\ell^2\,\sqrt{1-8\,\pi\,\eta^2\,\xi}}{16\,\beta^2}+\frac{M\,\ell\,(3\,\pi-4)\,\sqrt{1-8\,\pi\,\eta^2\,\xi}}{4\,\beta^2}.\label{c8}
\end{equation}

In Eq. (\ref{c8}), the first term arises due to the global monopoles (ordinary or phantom), the second term is due to the Schwarzschild black hole mass, the third term originates from the holonomy correction, while the subsequent terms reflect contributions from the holonomy correction, global monopoles, and black hole mass.

Now, we discuss a special case corresponds to the limit $\xi=0$. In this situation, the deflection angle from Eq. (\ref{c8}) reduces to
\begin{equation}
    \delta\phi \simeq \frac{4\,M}{\beta}+\frac{\ell}{\beta}+\frac{3\,\pi\,\ell^2}{16\,\beta^2}+\frac{M\,\ell\,(3\,\pi-4)}{4\,\beta^2}.\label{c88}
\end{equation}

Equation (\ref{c88}) is similar to the results obtained in the weak field limit for a holonomy-corrected Schwarzschild black hole without global monopoles \cite{PRD2}. Thus, we conclude that our result (\ref{c8}) is a generalization of the earlier results \cite{PRD2}, with the inclusion of global monopoles (ordinary or phantom) causing an increase in the deflection angle in the weak field limit.

For black hole space-time with an ordinary global monopole, $\xi=1$, the deflection angle is given by 
\begin{equation}
    \delta\phi \simeq \Big(\frac{1}{\gamma}-1\Big)\,\pi+\frac{4\,M}{\beta}+\frac{\ell}{\beta}+\frac{3\,\pi\,\ell^2\,\gamma}{16\,\beta^2}+\frac{M\,\ell\,(3\,\pi-4)\,\gamma}{4\,\beta^2},\quad \gamma=\sqrt{1-8\,\pi\,\eta^2}.\label{c12}
\end{equation}

For black hole space-time with phantom global monopole, $\xi=-1$, the deflection angle is given by 
\begin{equation}
    \delta\phi \simeq \Big(\frac{1}{\zeta}-1\Big)\,\pi+\frac{4\,M}{\beta}+\frac{\ell}{\beta}+\frac{3\,\pi\,\ell^2\,\zeta}{16\,\beta^2}+\frac{M\,\ell\,(3\,\pi-4)\,\zeta}{4\,\beta^2},\quad \zeta=\sqrt{1+8\,\pi\,\eta^2}.\label{c13}
\end{equation}

In the expression for the deflection angle, Eq. (\ref{c8}), the first term arises from either ordinary or phantom global monopoles, the second term is attributed to the black hole mass, the third term arises from holonomy corrections, and the last two terms represent a mixture of holonomy corrections and global monopoles.

\begin{center}
\begin{figure}[ht!]
\begin{centering}
\subfloat[$L=1$]{\centering{}\includegraphics[scale=0.72]{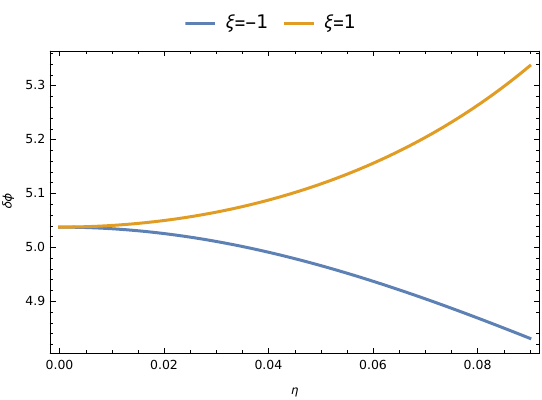}}\quad\quad\quad
\subfloat[$L=2$]{\centering{}\includegraphics[scale=0.72]{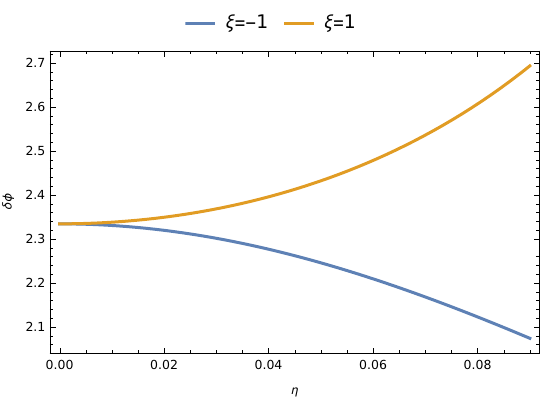}}\\
\subfloat[$L=3$]{\centering{}\includegraphics[scale=0.72]{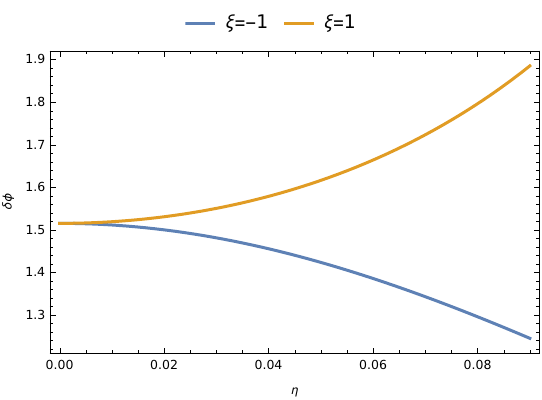}}\quad\quad\quad
\subfloat[$L=4$]{\centering{}\includegraphics[scale=0.72]{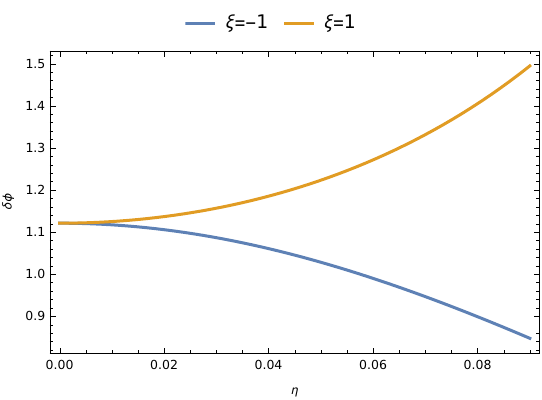}}
\caption{The deflection angle of photon light in weak field limit. Here, $M=1$ and $\ell=0.3$.}
\label{fig: 3}
\end{centering}
\end{figure}
\par\end{center}

Figure \ref{fig: 3} illustrates the deflection angle of photon light as a function of the energy scale $\eta$ for various values of the impact parameter $\beta=1, 2, 3, 4$. It is observed that as the impact parameter values increase, the corresponding deflection angle values decrease.

Now, in the limit $\ell \to 0$, indicating the absence of holonomy corrections, we derive the deflection angle of photon light for a black hole space-time with phantom global monopoles (ordinary and phantom) in the weak field limit. From Eq. (\ref{c6}) and using (\ref{c2}), we obtain
\begin{eqnarray}
    \Delta\phi=2\,\int^{u_0}_{0}\,\frac{du}{\sqrt{\alpha^2\,\Big[u^2_{0}-2\,M\,u^3_{0}-u^2+2\,M\,u^3\Big]}}. \label{c10}
\end{eqnarray}
Taking an approximation where $M <<1$, the expression for angle of deflection is given by
\begin{equation}
    \delta\phi \simeq \Big(\frac{1}{\sqrt{1-8\,\pi\,\eta^2\,\xi}}-1\Big)\,\pi+\frac{4\,M}{\beta}.\label{c9}
\end{equation}

Here, we see that the first term arises due to ordinary or phantom global monopoles ($\xi=\pm\,1$), and the second term from Schwarzschild BH mass. In the limit $\xi=0$, one will find the results similar to the Schwarzschild BH without any corrections. Thus, It is found that the deflection angle for BH with ordinary/phantom global monopole is greater/smaller than that of Schwarzschild BH.

\section{Conclusions}

Numerous studies have examined the phenomenon of photon deflection in diverse curved spacetime backgrounds, including those generated by black holes, wormholes, and topological defects. The present investigation focuses on examining the deflection of photon rays and geodesic motions of test particles in the background of holonomy corrected BH with phantom global monopoles. By employing the Lagrangian method, we have derived a one-dimensional energy expression and analyzed the effective potential for both null and time-like geodesics. Our findings demonstrate that the effective potential of the system is influenced by the phantom ($\xi=-1$) and ordinary global monopoles ($\xi=1$). We have generated a few graphs illustrating the effective potential for null and time-like geodesics within the considered BH background, considering different values of the angular momentum $L=1, 2, 3, 4$. These graphs are presented in Figures \ref{fig: 1}--\ref{fig: 2}. Furthermore, we have analytically derived the deflection angle for photon light and obtained an expression (\ref{c8}) under the weak field limit. This expression provides insights into the relationship between the deflection angle and relevant parameters present in BH geometry. To illustrate this influence, we generated Figure \ref{fig: 3} showcasing the variation of the angle of deflection for different values of $L$. This Figure \ref{fig: 3} provide visual evidence of how phantom global monopoles impact the deflection of light in our study. The deflection angle for the BH with ordinary global monopole is monotonically increasing function of $\eta$ and larger than the Schwarzschild limit. For the phantom global monopole, the deflection angle continuously decreases and is less than the deflection angle of Schwarzschild as well as BH with ordinary global monopole. In our future work, we will focus on deflection of light in the strong field limit and lens equation in this BH geometry.

\section*{Acknowledgements}

F.A acknowledges the Inter-University Centre for Astronomy and Astrophysics (IUCAA), Pune, India, for granting a visiting associateship.

\end{document}